\documentclass[11pt,a4paper]{article}
\pdfoutput=1
\usepackage{jheppub}
\usepackage{amsmath,epsfig}
\usepackage{mathrsfs}
\usepackage{amssymb}
\usepackage{mathtools}
\usepackage{graphics}
\usepackage{axodraw}
\usepackage{cancel}
\usepackage{slashed}
\usepackage{url}
\usepackage{hyperref}

\usepackage{color}
\usepackage[normalem]{ulem}

\newcommand{\be}{\begin{equation}}
\newcommand{\ee}{\end{equation}}
\newcommand{\bea}{\begin{eqnarray}}
\newcommand{\eea}{\end{eqnarray}}

\title{The seesaw portal in testable models of neutrino masses}

\author[a,b]{A.~Caputo,}
\author[a,b]{ P. ~Hern\'andez,}
\author[b]{J. ~L\'opez-Pav\'on,}
\author[a]{J.~ Salvado.}

\affiliation[a]{Instituto de F\'{\i}sica Corpuscular, Universidad de Valencia and CSIC, 
 Edificio Institutos Investigaci\'on, Catedr\'atico Jos\'e Beltr\'an 2, 46980 Spain}
\affiliation[b]{Theory Division, CERN 1211 Geneve 23, Switzerland}
\abstract{A Standard Model extension with two Majorana neutrinos  can explain the measured neutrino masses and mixings, and also account for the matter-antimatter asymmetry
 in a region of parameter space that could be testable in future experiments. The testability of the model relies to some extent on its minimality.  In this paper we address the possibility that the model might be extended by extra generic new physics which we parametrize
in terms of a low-energy  effective theory. We consider the effects of the   operators of the lowest dimensionality, $d=5$, and evaluate the 
upper bounds on the coefficients so that the predictions of the minimal model are robust. One of the operators gives a new production mechanism for the heavy neutrinos at LHC via higgs decays. The higgs can decay to a pair of such neutrinos that, being long-lived, leave a powerful signal of two displaced vertices. We 
estimate the LHC reach to this process. }


\begin{document}

\maketitle
\section{Introduction}

The measured neutrino masses  and mixings provide a strong hint of physics beyond the Standard Model, and this new physics might involve
new weakly coupled fields at the electroweak scale. The simplest of these possibilities is an extension of the
Standard Model with just two singlet Majorana fermions. This model has been shown to provide a natural 
explanation of the matter-antimatter asymmetry of the Universe \cite{Asaka:2005pn,Shaposhnikov:2008pf,Asaka:2011wq,Canetti:2010aw, Besak:2012qm,Canetti:2012kh,Drewes:2012ma,Ghisoiu:2014ena,Garbrecht:2014bfa,Shuve:2014zua,Abada:2015rta,Hernandez:2015wna,Hernandez:2016kel,Drewes:2016jae,Ghiglieri:2017gjz} via neutrino oscillations \cite{Akhmedov:1998qx} if the two heavy singlets
have masses in the range $[1,10^2]$ GeV, a range that implies that these states could be produced and searched for in   
  beam dump experiments (for a recent review see \cite{Alekhin:2015byh}) and colliders (for a sample set of references see \cite{Ferrari:2000sp,Graesser:2007pc,delAguila:2008cj,delAguila:2008hw,BhupalDev:2012zg,Das:2014jxa,Blondel:2014bra,Abada:2014cca,Helo:2013esa,Antusch:2015mia,Izaguirre:2015pga,Gago:2015vma}). This opens the possibility of 
 making this baryogenesis scenario testable: the combination of measurements of the mixings and masses of the heavy
 neutrino states, together with the determination of the CP phase $\delta$ in future oscillation experiments 
 and the amplitude of neutrinoless double beta decay might lead to a quantitative prediction of the baryon asymmetry in the universe 
 \cite{Hernandez:2016kel}.

The constraint of the measured neutrino masses and mixings fixes to a large extent the flavour mixings
of the heavy neutrino states  \cite{Hernandez:2016kel}. In particular, it has been shown that the ratios of mixings 
to different lepton flavours are strongly correlated with the CP violating phases of the PMNS matrix \cite{Caputo:2016ojx}. 
One of the most drastic implications of this minimal model is the vanishing of the lightest neutrino mass. If evidence of a 
non-vanishing lightest neutrino mass would come from future beta-decay experiments or from cosmological measurements, 
the minimal model with two neutrinos would not be able to explain it and an extension would be required. 

The question we would like to address in this paper is the following. If there is some extra new physics at a higher scale (for example 
 more singlet states), how does this new physics would modify the predictions of the low-energy theory represented by the minimal model with the two singlet states.
In particular, we would like to understand if the strong correlations between the light and heavy neutrino masses and mixings
 \cite{Caputo:2016ojx} that underlie the predictivity and testability of this model could be preserved or under what conditions they might be. 

A model independent way of answering this question is by building the effective theory and analysing
what modifications on the correlations higher dimensional operators can induce. The list of higher dimensional 
operators in the SM up to dimension $d= 6$ is well-known \cite{Weinberg:1979sa,Buchmuller:1985jz,Grzadkowski:2010es} and has been studied extensively. 
Interestingly however in the extended theory with fermion singlets \cite{Graesser:2007yj,delAguila:2008ir,Aparici:2009fh}, there are more $d=5$  
operators than in the SM.  The  relatively light singlet states provide a new portal into BSM physics.
 In this paper we will restrict ourselves to the lowest dimensional operators of $d=5$, which are expected to be dominant. 

 The paper is organized as follows. In the section 2 we describe the minimal model and the extension by $d=5$ operators. In section 3 we consider the 
 different  constraints on the $d=5$ operators from neutrino masses and LHC and in section 4 we conclude.

\section{The seesaw effective theory}

At energies much smaller than the new physics scale, $\Lambda$, the theory is just a type I seesaw model \cite{Minkowski:1977sc, GellMann:1980vs,Yanagida:1979as,Mohapatra:1979ia} with two 
Majorana neutrinos in the GeV range, with the Lagrangian
\begin{eqnarray}
{\cal L}_{SS} = {\cal L}_{SM}- \sum_{\alpha,i} \bar L_\alpha Y^{\alpha i} \tilde\Phi N_{i} - \sum_{i,j=1}^2 {1\over 2} \overline{N^{c}_{i}} M_N^{ij} N_{j}+ h.c. \nonumber
\label{eq:lag}
\end{eqnarray}
The leading effects of the new physics should be well described by higher dimensional operators of $d=5$ that can be constructed in a gauge invariant way with the Standard Model fields and the heavy Majorana neutrinos. These have been classified in Refs.~ \cite{Graesser:2007yj,delAguila:2008ir,Aparici:2009fh}. There are three independent operators: 
\begin{eqnarray}
{\mathcal O}_W &=& \sum_{\alpha,\beta} {(\alpha_{W})_{\alpha\beta}
 \over \Lambda} \overline{ L}_\alpha \tilde{\Phi}  \Phi^\dagger L^c_\beta +h.c., \\
{\mathcal O}_{N\Phi} &=& \sum_{i,j} 
{(\alpha_{N\Phi})_{ ij}\over \Lambda} \overline{N}_i {N}^c_{j} \Phi^\dagger \Phi
+h.c.,\\
{\mathcal O}_{NB} &=& \sum_{i \neq j}{(\alpha_{NB})_{ij} \over \Lambda} \overline{ N}_i
 \sigma_{\mu\nu} N_j^c B_{\mu\nu} + h.c. 
\end{eqnarray}
The first is the well known Weinberg operator ${\mathcal O}_W$ \cite{Weinberg:1979sa} that induces a new contribution to the light neutrino masses, independent of the contribution 
of the $N$ fields. The new operator ${\mathcal O}_{N\Phi}$ contributes to the $N$ Majorana masses, and interestingly gives additional couplings of these heavy neutrinos to the Higgs \cite{Graesser:2007yj,Graesser:2007pc}, which are not necessarily suppressed with the Yukawa couplings.   Constrains on this operator have been extensively studied in the context of Higgs portal dark matter \cite{Patt:2006fw}. 
In that case however, it is assumed that the Majorana fermion constitutes the dark matter and therefore does not decay, for which it is necessary to forbid the yukawa coupling to the lepton doublet. In our case the states can decay visibly in the detector. The last operator induces magnetic moments of the heavy neutrinos and the constraints have been studied in \cite{Aparici:2009fh}. 

We will see in the following that the direct constraints on the coefficients of these operators  are very different. It is therefore an important question whether large hierarchies could exist between the coefficients. On naturality grounds we would expect they should of the same order
\begin{eqnarray}
{\alpha_{W}\over \Lambda} \sim {\alpha_{N\Phi}\over \Lambda} \sim {\alpha_{NB}\over \Lambda}.
\label{eq:natural}
\end{eqnarray}
However this might not be the case if there exists an approximate global $U(1)_L$ symmetry \cite{Wyler:1982dd,Mohapatra:1986bd}. If the two Majorana neutrinos carry opposite lepton number charges, the Majorana mass term can be chosen to be of the form
\begin{eqnarray}
M_N = \left(\begin{array}{l l} 0 &M \\
M & 0 \end{array}\right),
\end{eqnarray}
to ensure an exact symmetry, implying degenerate heavy neutrinos and massless light neutrinos. We note that this approximate $U(1)_L$ symmetry is also usually invoked in the minimal model to justify yukawa couplings larger than the naive seesaw relation $Y \sim \sqrt{{M m_\nu\over v^2}}$, that are required to be within the sensitivity reach of future experiments.

 The operator ${\mathcal O}_{NB}$ is invariant, while the operator ${\mathcal O}_{N\Phi}$ is invariant provided the flavour structure of the coefficient is of the same form
 as the Majorana mass,
 \begin{eqnarray}
\alpha_{N\Phi} = \left(\begin{array}{l l} 0 & \alpha \\
\alpha & 0 \end{array}\right).
\end{eqnarray}
Only the Weinberg operator violates lepton number in this case.  If we assume the symmetry is approximately conserved, since we need to accommodate neutrino masses, it is technically natural in this context to assume the following hierarchy
\begin{eqnarray}
{\alpha_{W}\over \Lambda} \ll {\alpha_{N\Phi}\over \Lambda} \sim {\alpha_{NB}\over \Lambda}.
\end{eqnarray}
A hierarchy also arises in the context of minimal flavour violation \cite{Cirigliano:2005ck,Davidson:2006bd,Gavela:2009cd}. As discussed in \cite{Graesser:2007yj}, in this case the coefficients $\alpha_{W} \sim \alpha_{NB}\sim {\mathcal O}(Y^2)$, while there is no suppression of $\alpha_{N\Phi}$. 

\begin{figure}
\begin{center}
\includegraphics[width=0.6\columnwidth]{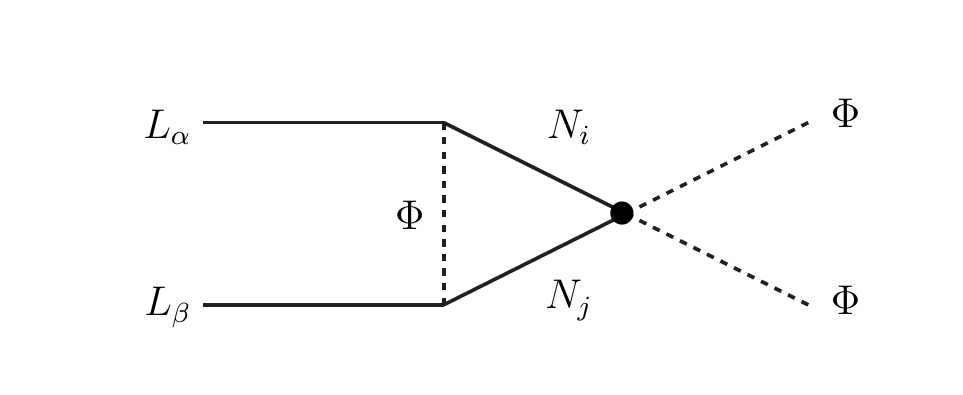} 
\end{center}
\caption{One loop contribution of ${\mathcal O}_{N\Phi}$ to the neutrino masses}
\label{fig:oneloop}
\end{figure}

On the other hand, if we drop completely the naturalness argument, we have to confront a hierarchy problem.  All the $d=5$ operators generically mix under renormalization. For example the diagram of Fig.~\ref{fig:oneloop} induces a contribution to the Weinberg operator at one loop of the form 
\begin{eqnarray}
\delta \left({\alpha_W\over \Lambda} \right) \propto \frac{1}{(4\pi)^2}\frac{Y \alpha_{N\Phi} Y^T}{4\Lambda}\log\frac{\mu^2}{M^2}.
\end{eqnarray}
Requiring this contribution to be of the same order or smaller than the tree level contribution implies the following bound
\begin{equation}
{\alpha_{N\phi}\over \Lambda} \lesssim {2\cdot 10^{13}\over \log\frac{\mu^2}{M^2}}
\left(\frac{10^{-6}}{\theta^2}\right)\left(\frac{{\rm GeV}}{M}\right)^2 {\alpha_W\over \Lambda},
\end{equation}
where $\theta$ is the mixing of the heavy states with the leptons.  This constraint is rather mild and leaves significant freedom to have a large hierarchy, even if no symmetry principle is assumed. 

\section{Constraints on the $d=5$ operators}

\subsection{Neutrino masses}

The operator ${\mathcal O}_{N\Phi}$  gives a correction to the Majorana neutrino mass matrix:
\begin{eqnarray}
M' = M + {v^2} {\alpha_{N\Phi}\over \Lambda}.
\end{eqnarray}
The light neutrino masses get a contribution from the two light states, and  from new physics at the scale $\Lambda$. We call this latter contribution $\delta m_\nu^W$:
\begin{eqnarray}
m_\nu &=& 
  \left(- {Y {1\over M'} Y^T}  + 2 {\alpha_W\over \Lambda}  \right) {v^2 \over 2} \equiv - {Y {1\over M'} Y^T}   {v^2 \over 2} + \delta m_\nu^W. 
\end{eqnarray}
The diagonalization of the complete light neutrino mass matrix gives:
\be
m_\nu=-\frac{v^2}{2}Y{1\over M'}Y^T + \delta m_\nu^W = U m_l U^{T}.
\label{eq:mnu}
\ee
where $U$ is the standard PMNS matrix, $m_l=\text{diag}\left(m_1,m_2,m_3\right)$ are the light neutrino masses. 
We know that in the limit $\delta m_\nu^W\rightarrow 0$ , 
 $m_1(m_3) \rightarrow 0$ for the normal (inverted) hierarchy (NH/IH) or equivalently the condition
%
\be
\det\left[Um_l U^{T} - \delta m_\nu^W\right]=0
\ee
must  hold.
Perturbing in $(\delta m_\nu^W)_{ij} = \delta_{ij}$  and the small light neutrino parameters $\theta_{13}\sim \theta_{23}-\pi/4 \sim r \equiv {\Delta m^2_{\rm sol}/\Delta m^2_{\rm atm}}$, all of which are taken of the same order,  $
{\mathcal O}(\epsilon)\sim {\mathcal O}(\delta)$, we obtain at leading order:
\bea
\text{NH:} & & \\
m_1&=&-\left[U^\dagger m_l U^{*}\right]_{11}
\nonumber\\
&=&\frac{-e^{2 i\phi_1}}{2}\left[2\delta_{11}c_{12}^2+
(\delta_{22}-2\delta_{23}+\delta_{33})s_{12}^2 +\sqrt{2}(\delta_{13}-\delta_{12})\sin2\theta_{12}\right]
+\mathcal{O}\left(\epsilon\, \delta \right),
\nonumber\\
\text{IH:} &&\\
m_3&=&-\left[U^\dagger m_l U^{*}\right]_{33}=\frac{-e^{2i \phi_1}}{2}\left(\delta_{22}
+2 \delta_{23}+\delta_{33}\right) +\mathcal{O}\left(\epsilon\, \delta \right).
\eea
 Barring fine-tuned cancellations among the $\delta_{ij}$ coefficients, we have therefore
\be
\delta_{ij} = \frac{(\alpha_{W})_{ij}v^2}{\Lambda}\sim {\mathcal O}(1) \,m_{1,3}.
\label{eq:dm}
\ee
%
The lightest neutrino mass is thus the measure of the non-standard contributions from ${\mathcal O}_W$.

\subsection{Heavy neutrino Mixing}

\begin{figure}
\begin{center}
 \includegraphics[width=0.5\columnwidth]{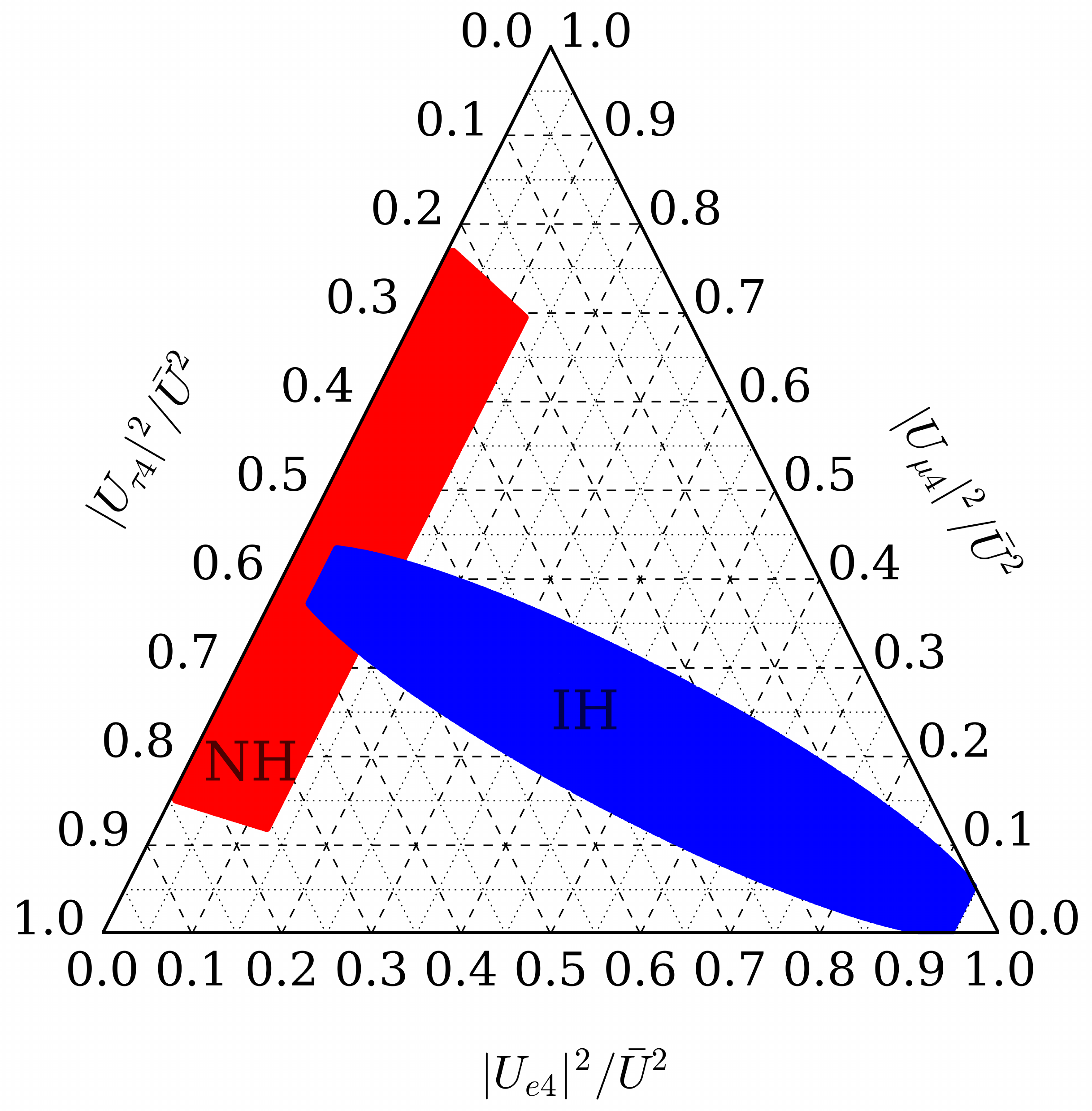} 
 \caption{Ternary diagram for the normalized flavour mixings  $|U_{\alpha i}|^2/\bar{U}^2$ (in the large mixing regime) fixing the known oscillation parameters to their best fit values and varying the CP phases from $[0, 2 \pi]$ for NH (red) and IH (blue).}
 \label{fig:mixings}
 \end{center}
\end{figure}

The heavy states mix with the charged leptons inducing charged and neutral currents as well as higgs interactions. All these get modified by the $d=5$ operators. 
In the mass basis the heavy neutrinos interact with the $Z$ and $W$ via the interactions:
\begin{eqnarray}
- {g \over \sqrt{8}} U_{\alpha 3+i} \bar{l}_\alpha \gamma_\mu (1 -\gamma_5) N_i W_\mu^-  - {g \over 4 \cos\theta_W} U_{\alpha 3+i}  \bar{\nu}^\alpha \gamma_\mu(1-\gamma_5) N_i  Z_\mu+ h.c. 
\end{eqnarray}
where $N_i$ are now the mass eigenstates, $i=1,2$ and $U_{\alpha 3+i}$ are their mixings to the flavours $\alpha=e, \mu, \tau$. These mixings are related to the Yukawa couplings 
as 
\begin{eqnarray}
U_{\alpha 3+i} \simeq {v\over \sqrt{2}} (Y 
M'^{-1})_{\alpha i}.
\end{eqnarray}
We can use the Casas-Ibarra trick \cite{Casas:2001sr} to parametrize $Y$ in terms of ${\widetilde m}$ and ${\widetilde U}$ where
\be
m_\nu -\delta m_\nu^W = -\frac{v^2}{2}YM^{-1}Y^T\equiv \widetilde{U}\widetilde{m} \widetilde{U}^{T},
\label{eq:mnu}
\ee
and a generic 2$\times$2 complex orthogonal matrix $R$. In this parametrization the mixings are
\be
U_{\alpha 3+i}=i (\widetilde{U}\widetilde{m}^{1/2}R^\dagger M'^{-1/2})_{\alpha i}. 
\label{eq:CI}
\ee
 In the limit $\delta m_\nu^W \rightarrow 0$, ${\widetilde m}$ and ${\widetilde U}$ coincide with the light neutrino masses
and mixing matrix. It has been shown that in this case the flavour structure of the mixings is approximately fixed those neutrino oscillation observables \cite{Gavela:2009cd,Ibarra:2010xw,Hernandez:2016kel}. Simple analytical approximations were worked out in \cite{Hernandez:2016kel} that should be accurate in the sensitivity region of future experiments. These are shown in 
 Fig.~\ref{fig:mixings}, where we show the
 allowed range for normalized flavour mixings: $|U_{\alpha 3+i}|^2/\bar{U}^2$, with $\bar{U}^2 \equiv \sum_\alpha |U_{\alpha 3+i}|^2$, fixing the known oscillation parameters to their best fit values \cite{Esteban:2016qun}, and  varying the CP violating phases $\delta$ and $\phi_1$ in the whole physical range  (the second Majorana phase is unphysical for two heavy  neutrinos). It has been shown that the determination of these ratios in future experiments would give very valuable information on the CP phases \cite{Caputo:2016ojx}. 

 In general  however the parameters  $\tilde U$ and $\tilde m$ are sensitive to new physics. We can relate them  to the physical mixings 
and masses using perturbation theory in $\delta m_\nu^W={\mathcal O}( \delta)$. For NH we have
\bea
\widetilde{U}\widetilde{m}^2 \widetilde{U}^\dagger &=& U\text{diag}\left(0,m_2^2,m_3^2\right)U^{T} + \left\{U\text{diag}\left(0,m_2,m_3\right)
U^{T},\delta {m_\nu^W}^\dagger\right\}
+\mathcal{O}\left(\delta
^2 \right),
\label{eq:mss.mssdag}
\eea
where $\{,\}$ is the anticommutator. 
For IH the result is the same with $m_3 \rightarrow m_1$ and moved to the first position. 
Here we have used that $m_1/m_3 = {\mathcal O}(\delta)$. The first order corrections  $\mathcal{O}\left(\delta \right)$ to $\tilde m$ and $\tilde U$ are presented in appendix A.

\begin{figure}
\begin{center}
\includegraphics[width=0.4\columnwidth]{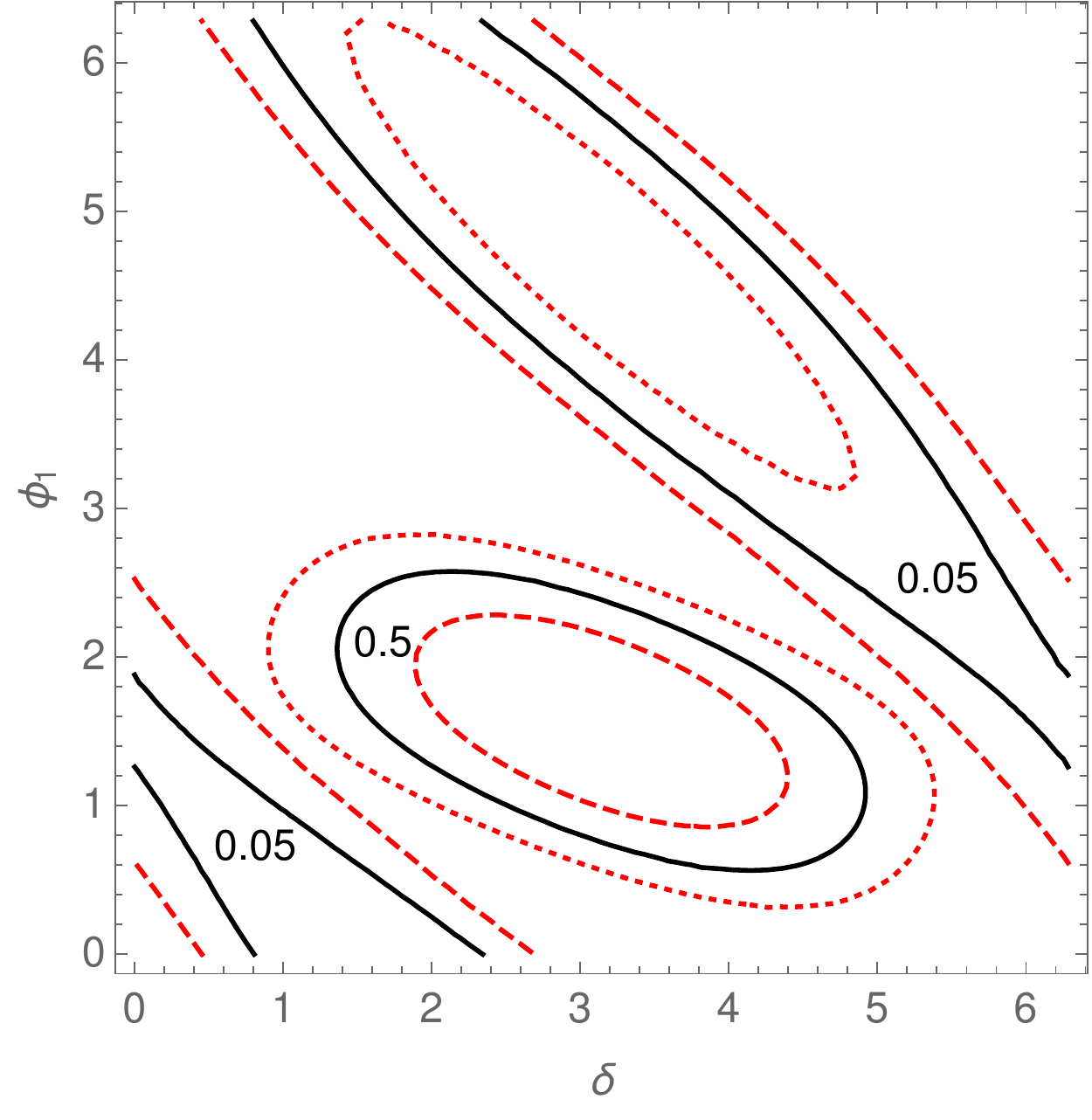}
 \includegraphics[width=0.4\columnwidth]{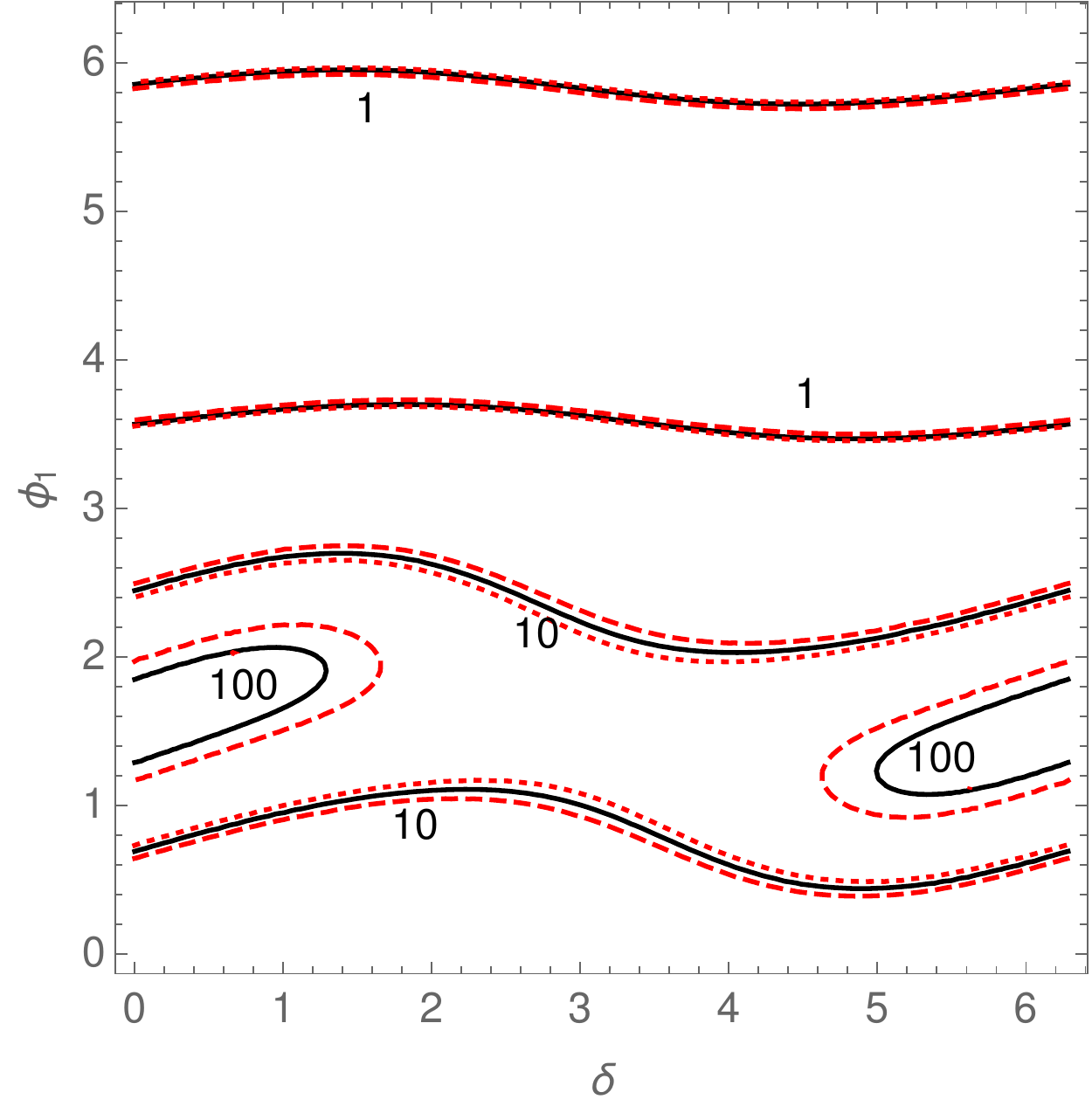} 
  \caption{Contours  of constant ratio $|U_{ei}|^2/|U_{\mu i}|^2$ as a function of the two CP phases $\delta, \phi_1$ for NH(left) and IH(right). The three curves correspond to 
  no $\Lambda^{-1}$ corrections (solid black) and minimal or maximal (red dashed) deviations within the range $c_\alpha, c'_\alpha \in [-1,1]$.  }
 \label{fig}
 \end{center}
\end{figure}

Including these in eq.~(\ref{eq:CI}) we find that the approximate expressions for the mixings in the absence of extra dynamics derived in ref.~\cite{Hernandez:2016kel} get modified by corrections $\delta |U_{\alpha i}|^2$.  For  NH we get
\bea
\delta|U_{e4}|^2 M_1 = 
A \,c_e\frac{m_1}{\sqrt{\Delta m^2_{atm}}},\;\;\;
\delta|U_{\mu 4}|^2 M_1 =
A \,c_\mu\frac{m_1}{\sqrt{r \Delta m^2_{atm}}},
\eea
where $A=e^{2\gamma}\sqrt{\Delta m^2_{atm}}/4$. Similarly for IH:
\bea
\delta |U_{e4}|^2 M_1 =  A \,c'_e\frac{m_3}{\sqrt{\Delta m^2_{atm}}}, \;\;\;
\delta |U_{\mu 4}|^2 M_1 =  A \,c'_\mu\frac{m_3}{\sqrt{\Delta m^2_{atm}}},
\eea
 and $\delta|U_{\alpha 4}|^2 M_1 = \delta|U_{\alpha 5}|^2 M_2$ at this order for both hierarchies.
Barring fine tuning among the different entries in $\delta m_\nu^W$ and taking into account Eq.~(\ref{eq:dm}), the
coefficients $c_\alpha$ and $c_\alpha'$ are expected to be $\mathcal{O}\left(1 \right)$. In Fig.~\ref{fig} we show the impact of these corrections  to the ratio $|U_{e4}|^2/|U_{\mu 4}|^2$ dependence on the CP phases for
$m_{1,3}=0.1 \sqrt{\Delta m^2_{sol}}$ and varying $c_\alpha$ and $c_\alpha'$ between $-1$ and $1$ to maximize the difference.
The corrections to the flavour ratios are therefore of $O(m_{\rm lightest}/\sqrt{\Delta m^2_{atm}})$ for IH and of $O(m_{\rm lightest}/\sqrt{\Delta m^2_{solar}})$ for NH. It will be hard
experimentally to ensure that these corrections are sufficiently small,  in other words to prove that $m_{\rm lightest}$ is significantly smaller than $\sqrt{\Delta m^2_{\rm atm}}$ for IH and 
$\sqrt{\Delta m^2_{\rm solar}}$ for NH. On the
other hand, measuring a lightest neutrino mass above this value will surely imply that corrections to the predictions of the minimal model are likely to be significant. Alternatively finding 
the flavour ratios in the regions given by Fig.~\ref{fig:mixings} would be a strong indication that only two Majorana states give the dominant contribution to the light neutrino masses 
and therefore the  lightest neutrino mass is significantly smaller than the mass splittings.

\subsection{Higgs-Neutrino Interactions}

\begin{figure}
\begin{center}
\includegraphics[width=0.4\columnwidth]{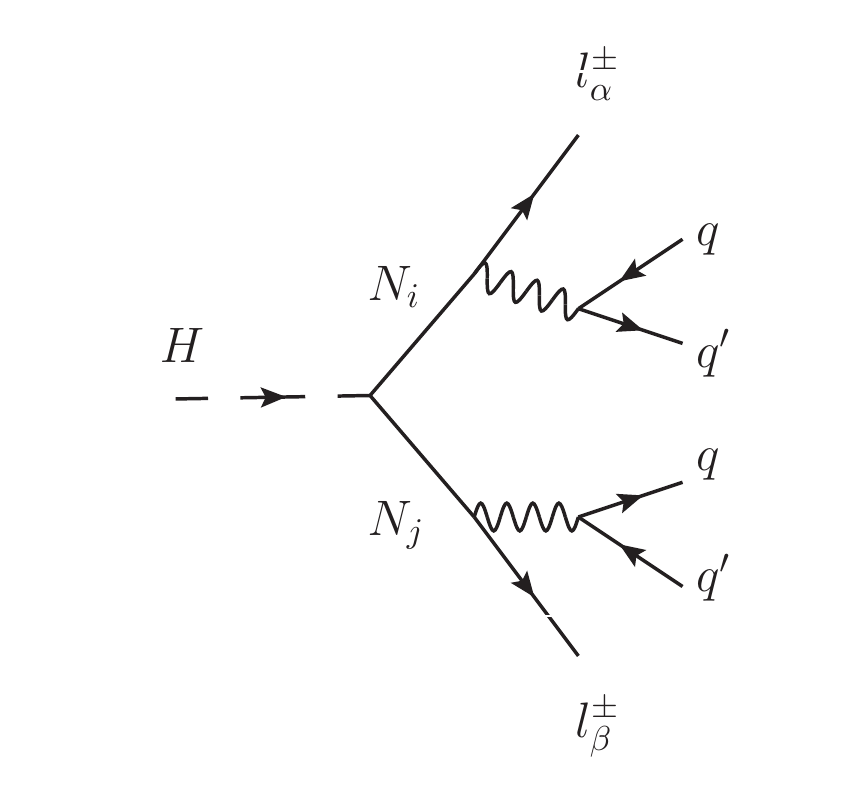}
 \caption{Higgs decay to two heavy neutrinos leading to displaced vertices}
 \label{fig:diagram}
\end{center}
\end{figure}

Interestingly the operator ${\mathcal O}_{N\Phi}$ induces new higgs interactions with the right handed neutrinos. In particular for $M_i \leq {M_H\over 2}$, the Higgs can decay 
to two heavy neutrinos via this coupling
\begin{eqnarray}
{\mathcal L} \supset -{v \over \sqrt{2} \Lambda}  
H \overline{N}^c \alpha_{N\Phi} N + h.c.
\label{eq:hdecay}
\end{eqnarray}
This decay leads to a spectacular signal at LHC, which is that of a pair of displaced vertices (DVs) since
the $N$ decay only via mixing and they have a long decay length. The process is depicted in Fig.~\ref{fig:diagram} Note that compared to the minimal model, where the sterile neutrinos are singly produced via mixing, in this case the production is controlled by the coefficient $\alpha_{N\Phi}$ and therefore one should have access to much smaller mixings, as long as they lead to lifetimes are within reach 
of DV searches within the detectors. 

Our aim in this section is to do a simple estimate of
 the bounds on the coupling $\alpha_{N\Phi}/\Lambda$  from searches of higgs decays
to two displaced vertices at LHC. A closely related calculation has been done in the context of $U(1)'$ models in \cite{Accomando:2016rpc}, where the signal selection has been 
performed following recent  searches by the CMS collaboration \cite{CMS:2014hka,CMS:2014muon}. We have considered two different analyses: 1) a search of displaced tracks in the inner tracker where at least
one displaced lepton, $e$ or $\mu$, is reconstructed from each vertex; 2) a search for displaced tracks in the muon chambers and outside
the inner tracker where at least one $\mu$ is reconstructed from each vertex.  The charges are not restricted and therefore events with same-sign or opposite sign leptons 
are possible. 

For simplicity we will consider only semileptonic decays of the $N_i$ which give rise to two lepton final states through the decay
\begin{eqnarray}
N_i \rightarrow  l^\pm W^\mp \rightarrow l^\pm q \bar{q}'.
\end{eqnarray}

\begin{figure}
\begin{center}
 \includegraphics[width=0.6\columnwidth]{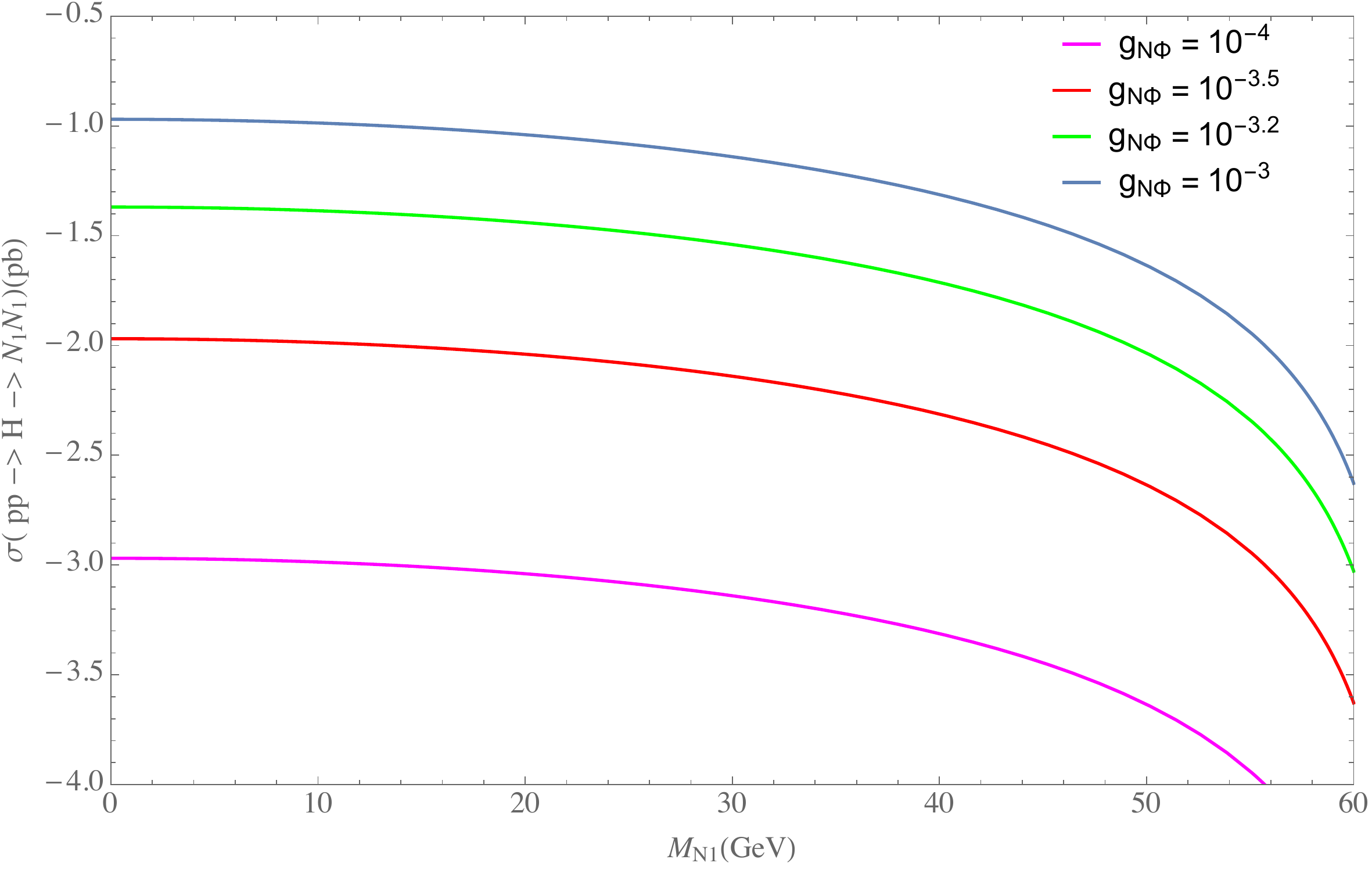} 
\caption{Cross section $pp\rightarrow H \rightarrow N_1 N_1$ in pb as a function of the heavy neutrino mass, $M_1$ for a few values of $g_{N\Phi} \equiv {v (\alpha_{N \Phi})_{11} \over \sqrt{2} \Lambda}$.}
 \label{fig:prod}
\end{center}
\end{figure}

\begin{figure}
\begin{center}
 \includegraphics[width=0.6\columnwidth]{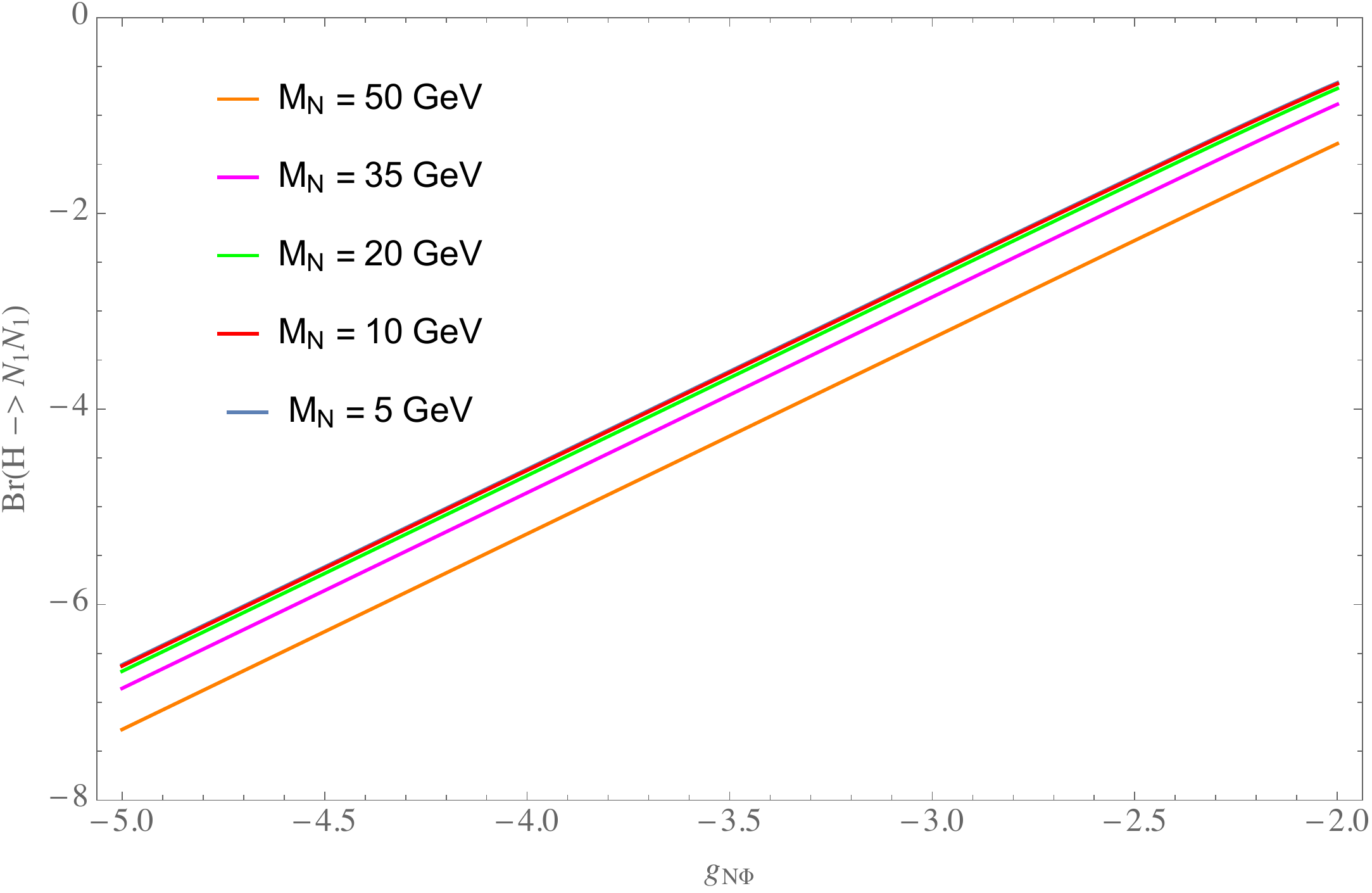} 
\caption{Br$(H\rightarrow N_1 N_1)$ as a function of the $g_{N\Phi}$ for various masses $M_1$. }
 \label{fig:br}
\end{center}
\end{figure}

We consider a parton-level Monte Carlo analysis using Madgraph5 \cite{Alwall:2014hca} at LHC with a center-of-mass energy  of 13TeV and 300 fb$^{-1}$ luminosity. We include only the dominant gluon fusion higgs production and we consider the production of just one neutrino species, $N_1$. The production cross section $pp\rightarrow h \rightarrow N_1 N_1$ is shown in Fig.~\ref{fig:prod} as a function of the heavy neutrino mass for various values of the 
 coupling $g_{N\Phi} \equiv {v (\alpha_{N \Phi})_{11} \over \sqrt{2} \Lambda}$. In Fig.~\ref{fig:br} we show the Br($H \rightarrow N_1 N_1)$ as a function of $g_{N\Phi}$ for various values of the mass (here we assume the higgs decays just to one neutrino). 

\begin{figure}
\begin{center}
 \includegraphics[width=0.6
 \columnwidth]{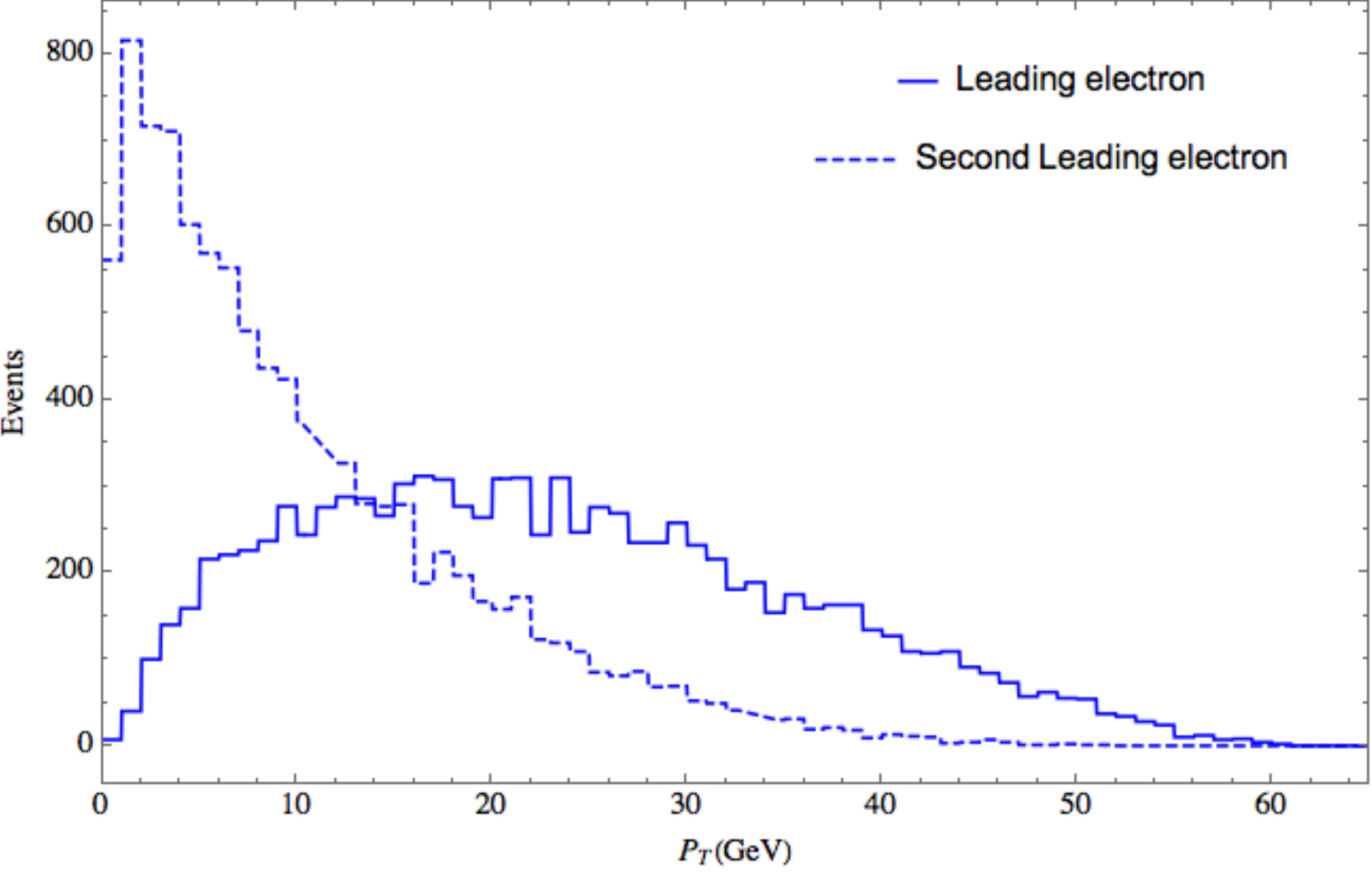}  
 \caption{$p_T$ distribution of the two leading electrons for $M=15$GeV.}
 \label{fig:pt}
\end{center}
\end{figure}

 The $p_T$ of the two leading leptons is shown in Fig.~(\ref{fig:pt}). Following \cite{Accomando:2016rpc}, the signal selection is done  by requiring two  lepton tracks, $e$ or $\mu$ that satisfy the following kinematical cuts on transverse momentum, pseudorapidity and isolation of the two tracks: 
\begin{eqnarray}
p_T(l) > 26~ {\rm GeV}, \;\; |\eta| < 2,\;\;\; \Delta R > 0.2,\;\;\; \cos \theta_{\mu\mu} > -0.75. 
 \end{eqnarray}
 In the case of muons a constraint in the opening angle $\theta_{\mu\mu}$ is imposed in order to reduce the cosmic muon background. 
The efficiencies resulting from these consecutive cuts for various neutrino masses are shown in Tables~\ref{tab:kin1} and \ref{tab:kin2}. Note that they do not depend on the mixings $|U_{\alpha i}|^2$. 
\begin{table}
\begin{center}
\begin{tabular}{|c|c|c|c|c|}
 \hline
$ee$  
 & $M_1$ = 10GeV & $M_1= 20$GeV& $M_1=30$GeV & $M_1=40$GeV\\
 \hline
 \hline
 $p_T$  & 6.4$\%$ & 7.0$\%$ & 5.6$\%$ & 4.5$\%$\\
 \hline
 $\eta$ & 4.2$\%$ & 4.8$\%$ & 4$\%$ & 2.9$\%$\\
 \hline
 $\Delta R$  & 4.2$\%$ & 4.8$\%$ & 4$\%$ & 2.9$\%$\\
  \hline
 \end{tabular}
\caption{Signal efficiciencies after consecutive cuts on $p_T$, $\eta$ and 
$\Delta R$ for the $ee$ channel in the inner tracker, for various heavy neutrino masses.}
\label{tab:kin1}
  \end{center}
\end{table} 

\begin{table}
\begin{center}
\begin{tabular}{|c|c|c|c|c|}
 \hline
 $\mu\mu$& $M_1=10$GeV & $M_1=20$GeV& $M_1=30$GeV & $M_1=40$GeV\\
 \hline
 \hline
 $p_T$  & 7.0$\%$ & 6.8$\%$ & 6.0$\%$ & 4.7
 $\%$\\
 \hline
 $\eta$ & 4.7$\%$ & 4.9$\%$ & 4$\%$ & 3.2$\%$\\
 \hline
 $\Delta R$  & 4.7$\%$ & 4.9$\%$ & 4$\%$ & 3.2$\%$\\
  \hline
   $\cos\theta_{\mu\mu} $  & 3.2$\%$ & 3.6$\%$ & 3.0$\%$ & 2.7$\%$\\
   \hline
 \end{tabular}
\caption{Signal efficiciencies after consecutive cuts on $p_T$, $\eta$ and 
$\Delta R$ for the $\mu\mu$ channel in the muon chamber for various heavy neutrino masses.}
\label{tab:kin2}
  \end{center}
\end{table} 

For each event the decay length of each neutrino in the laboratory frame is obtained from their simulated momenta and the distance travelled, $L$, is randomly sampled according to the expected exponential distribution.  The impact parameter of the lepton in the transverse direction $d_0$  is defined as
\begin{eqnarray}
d_0 = | L_x p_y - L_y p_x|/p_T,
\end{eqnarray}
where $L_{x,y,z}$ are the $X,Y,Z$ projections of the vector $L \vec{u}$, where $\vec{u}$ is the velocity of the heavy neutrino, and 
$(p_x,p_y)$ is the momentum of the lepton in the transverse direction. 

\begin{figure}
\begin{center}
 \includegraphics[width=0.6\columnwidth]{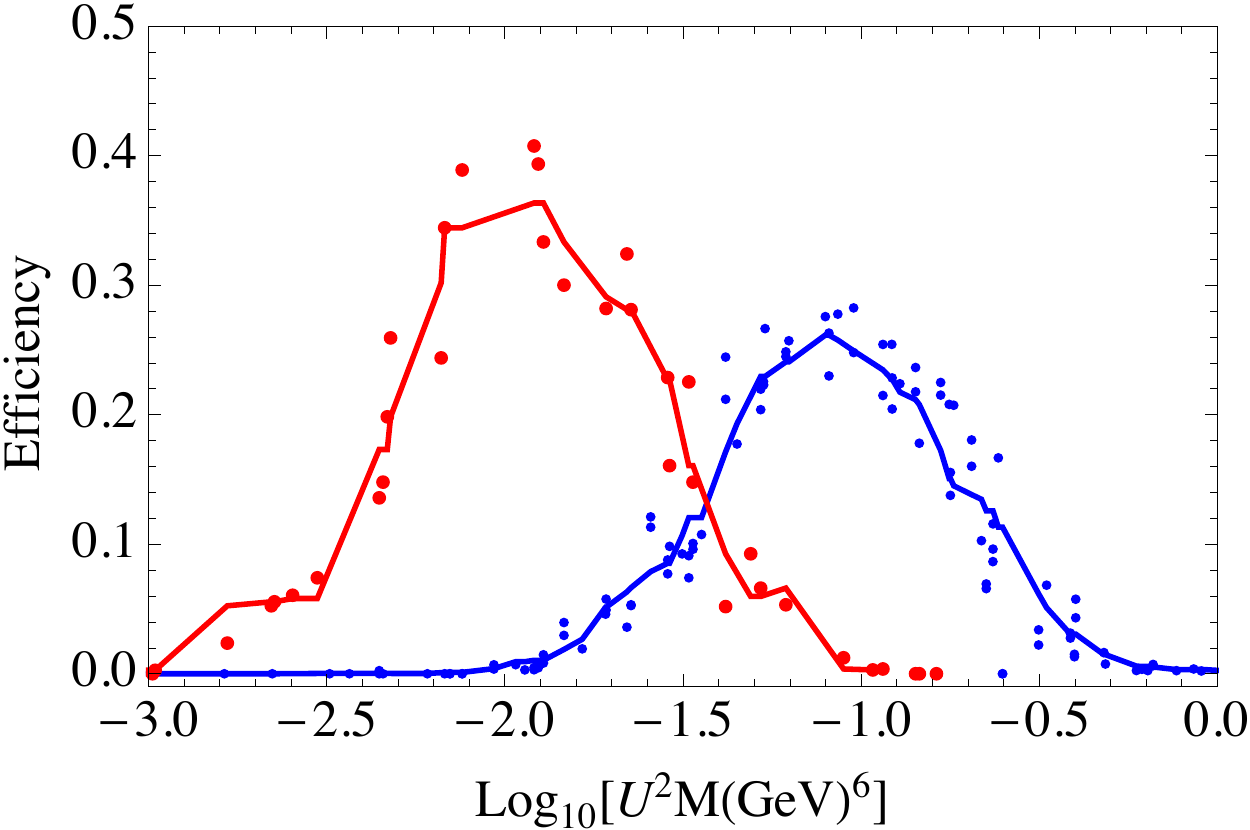}
 \caption{Acceptance of the cuts in eqs.~(\ref{eq:tracker}) in blue and (\ref{eq:chamber}) in red as a function of the combination $U^2 M^6$. }
 \label{fig:dis}
\end{center}
\end{figure}

The cuts associated to the displaced tracks are different for the two analyses \cite{Accomando:2016rpc}:
\begin{itemize}
\item Inner tracker (IT): 
 \begin{eqnarray}
 10 {\rm cm} < |L_{xy}|< 50 {\rm cm},\;\; |L_z| \leq 1.4 {\rm m},\;\;
  d_0/\sigma_d^t > 12,
  \label{eq:tracker}
 \end{eqnarray}
where $\sigma_d^{t}\simeq 20 \mu$m is the resolution in the tracker.
\item Muon chambers (MC): 
 \begin{eqnarray}
 |L_{xy}|\leq 5 {\rm m},\;\; |L_z| \leq 8 m, \;\; d_0/\sigma_d^{\mu} > 4,
  \label{eq:chamber}
   \end{eqnarray}
where the impact parameter resolution in the chambers is $ \sigma_d^{\mu}\sim 2$cm.  
\end{itemize}
Since $\langle L^{-1} \rangle \propto U^2 M^6$, we expect that these cuts depend mostly on this combination of mass and mixing with some dispersion. In Figs.~\ref{fig:dis} we show the acceptance given by 
cuts of eqs.~(\ref{eq:tracker}) and ~(\ref{eq:chamber}) as a function of $U^2 M^6$.  As expected the two analyses are sensitive to different decay lengths. 
 We have checked that the global efficiencies resulting for each of these cuts agree reasonably well with the results quoted in \cite{Accomando:2016rpc}.  

Concerning the flavour structures, we have considered two scenarios:

$\bullet$ Scenario A: only the decay of the higgs to one neutrino species, $N_1$, is considered. Obviously if the other state has a mass below $M_H/2$, it will also contribute via the decays
$H \rightarrow N_2 N_2$ and $H\rightarrow N_1 N_2$. These will be sensitive however to different entries of the matrix $\alpha_{N\Phi}$. 

\begin{figure}
\begin{center}
 \includegraphics[width=0.7\columnwidth]{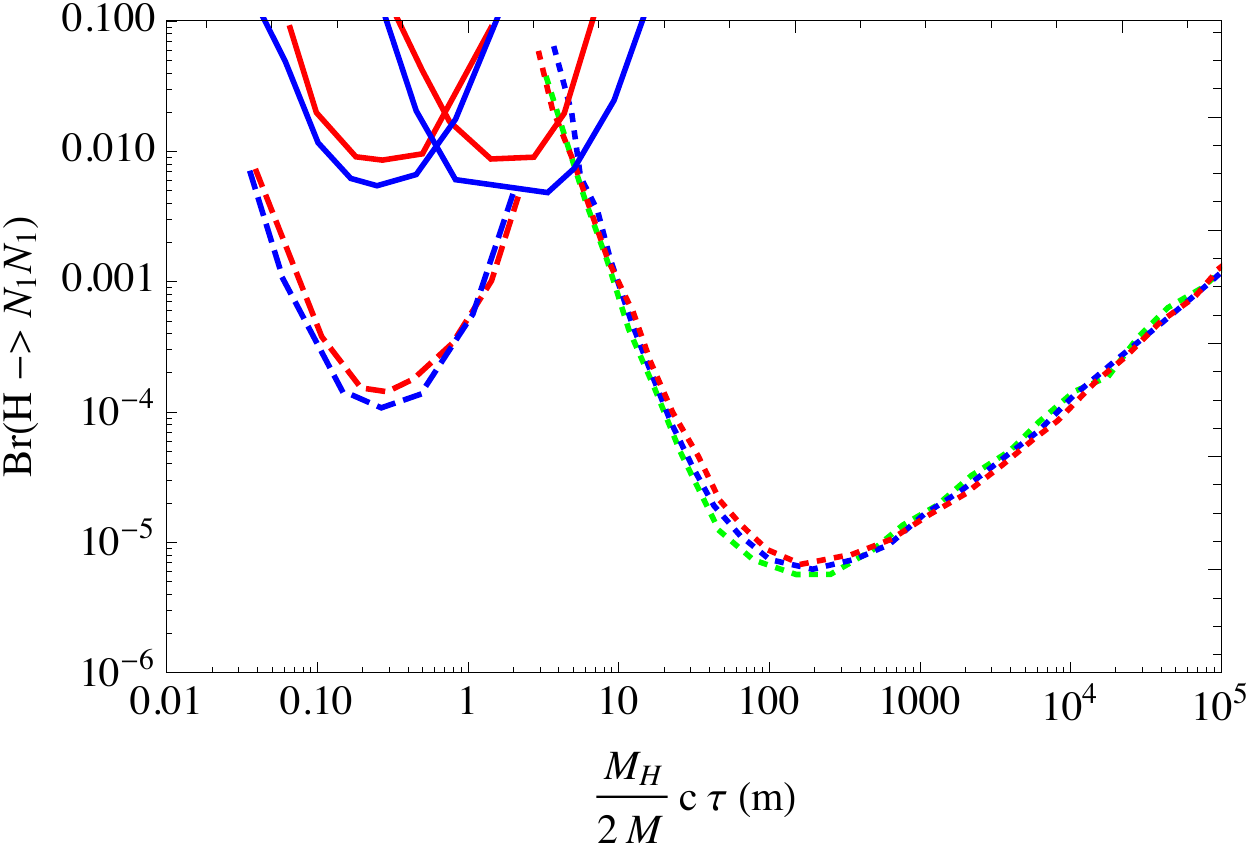} 
 \caption{Expected limits  at LHC (13 TeV, 300 fb$
^{-1}$) on the $BR(H\rightarrow N_1 N_1)$ for scenario A as a function of ${M_H\over 2 M} c \tau$ in meters. The solid lines correspond to the IT and MC 
 analyses with $M_1=20$GeV (blue) and $M_1=35$GeV (red) for NH and CP phases $\delta=0, \phi_1=\pi/2$. The dashed lines are for IH and the same CP phases for IT/MC
. The dotted lines correspond to the estimated reach of Mathusla in the HL-LHC (14TeV, 3000 fb$^{-1}$) \cite{Chou:2016lxi} for masses $5, 20, 40$GeV, assuming 
  that all the decays of $N_1$ are observable.    }
 \label{fig:brlimit}
 \end{center}
\end{figure}

 In Fig.~\ref{fig:brlimit} we show the  $BR(H\rightarrow N_1 N_1)$ corresponding to 4 signal events as a function of ${M_H \over 2 M} \times
 c \tau$ which is  approximately the decay length in the laboratory frame. We present the results that include the events producing two displaced electrons in the inner tracker, and either 
 two displaced  $\mu$ in the inner tracker (IT analysis) or  two displaced muons in  the muon chamber (MC analysis). We compare these limits with the those  of the Mathusla project of the HL-LHC with 3000$fb^{-1}$ as obtained in ref.~\cite{Chou:2016lxi}, where all the decays of the $N_1$ are assumed observed. The limit this implies on the coefficient of the operator, $g_{N\Phi}$ can then be read from Fig.~\ref{fig:br}.
 
\begin{figure}
\begin{center}
 \includegraphics[width=0.7\columnwidth]{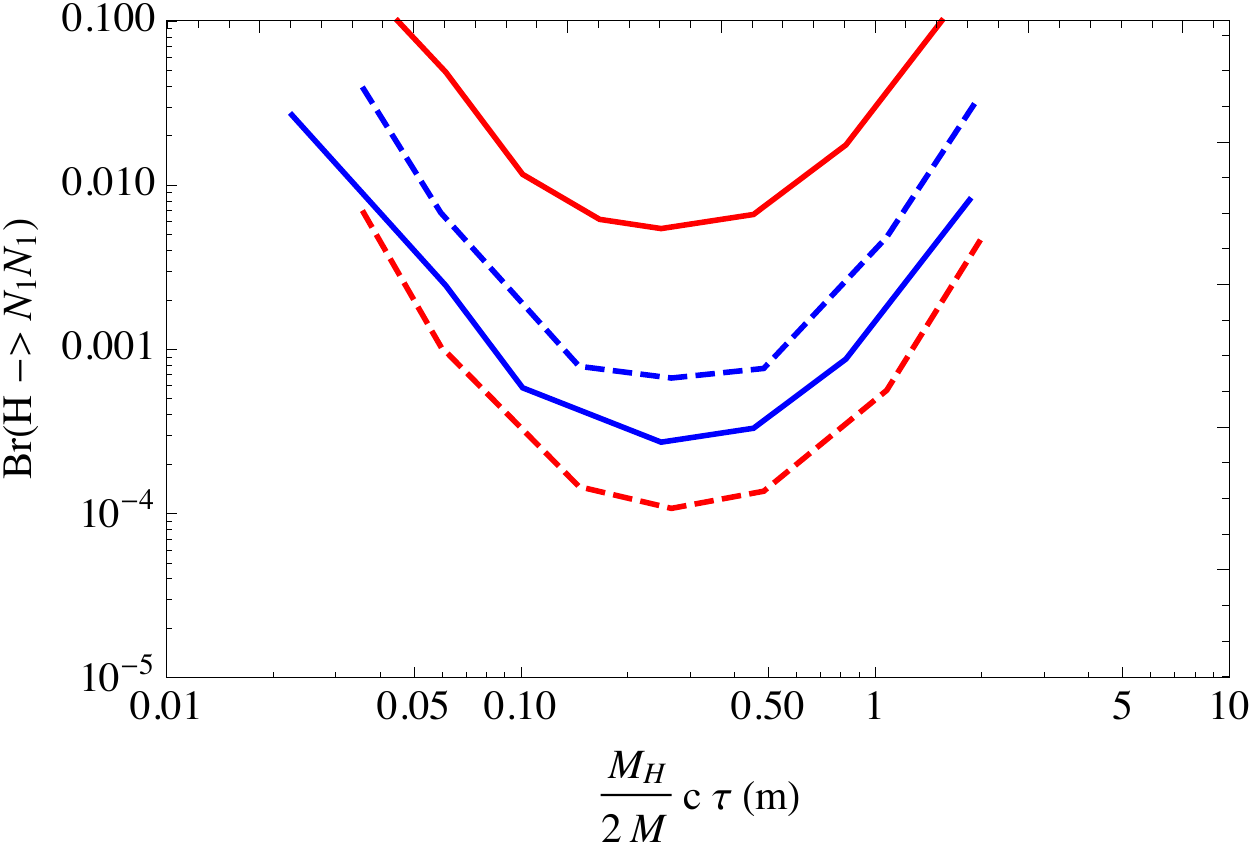} 
 \caption{Expected limits  at LHC (13 TeV, 300 fb$
^{-1}$) on the $BR(H\rightarrow N_1 N_1)$ for scenario A as a function of ${M_H\over 2 M} c \tau$ in meters. The mass is $20$ GeV and the CP phase $\phi_1=-\pi/2$ (blue) or 
$\phi_1=\pi/2$ (red) for NH (solid) and IH (dashed). }
 \label{fig:brphases}
 \end{center}
\end{figure}

 As we have seen the flavour structure of the mixings depends strongly on the hierarchy and the CP phases (see Fig.~\ref{fig:mixings}). We fix the CP phases to $\delta=0,\phi_1=\pi/2$. For this choice and NH, the muon channels dominates, while for IH it is the electron that dominates. In the former case
therefore the reach of the two analysis IT and MC are quite different, showing the reach in complementary regions of $c \tau$, while  for the latter most of the sensitivity comes from the electron search which contributes the same to both analyses as it can only be done with the tracker. The dependence on the phases is illustrated in Fig.~\ref{fig:brphases} where the phase $\phi_1$ is chosen approximately to maximize/minimize the sensitivity.

$\bullet$ Scenario B: the flavour structure of the matrix $\alpha_{N\Phi}$ is such that the operator is symmetric under a global $U(1)_L$ symmetry.  The symmetry is very transparent in the basis where the Majorana neutrino mass matrix and $\alpha_{N\Phi}$ are of the form
\begin{eqnarray}
M = \left(\begin{array}{l l} 0 &M \\
M & 0 \end{array}\right),
\alpha_{N\Phi} = \left(\begin{array}{l l} 0 & \alpha \\
\alpha & 0 \end{array}\right).
\end{eqnarray}
This pattern ensures that a lepton number is preserved 
under which $N_1$ and $N_2$ have opposite charges. In the same basis only $N_1$ has Yukawa couplings.  As a result of the approximate symmetry  same sign lepton decays vanish, the massive state is a Dirac fermion. The branching ratios are however unchanged.  The bounds on this scenario also can be read from Fig.~\ref{fig:brlimit}, if we interpret the $y$-axis as the total branching ratio of the higgs to the  heavy Dirac state.  On other hand the dependence of this branching ratio on the coupling $g_{N\Phi}$ is twice larger as that shown in Fig.~\ref{fig:br}.

\section{Discussion}

In order to explain the observed neutrino masses in the context of Type I seesaw models, only two heavy Majorana singlets are required. This is the minimal extension of the Standard Model that can accommodate neutrino masses. The flavour structure of the mixings of the heavy neutrino mass eigenstates is strongly correlated with the light neutrino masses and the PMNS matrix. If the mass of these heavy states is in the electroweak range, they can be searched for in future experiments within the high intensity frontier. A putative measurement of their masses and mixings  might demonstrate the origin of neutrino masses. The predictivity of this minimal model 
relies on the assumption that only these two states give the dominant contribution to neutrino masses. Any extension of this minimal scenario can modify these predictions. 
We have considered the possible modifications on the minimal scenario induced by generic new physics at some higher scale $\Lambda$. Three operators can parametrize these modifications at leading order in $\Lambda^{-1}$. 

Assuming the coefficients of these operators are all of the same order, the strongest bound comes from the bound on the lightest neutrino mass:
\begin{eqnarray}
\left|{\alpha_{W} v^2 \over 2 \Lambda}\right| \leq {\mathcal O}(1) m_{\rm lightest} \rightarrow {\alpha_{W} \over \Lambda} \leq 6 \times 10^
{-9} {\rm TeV}^{-1} \left({m_{\rm lightest} \over 0.2 {\rm eV} }\right)
\end{eqnarray}
A more stringent bound needs to be imposed generically to preserve the flavour predictions of the minimal model in the presence of new physics. Corrections to these predictions from the Weinberg operator are of ${\mathcal O}\left({m_{\rm light} \over \sqrt{\Delta m^2 _{\rm atm}}}\right)$ for IH or ${\mathcal O}\left({m_{\rm light} \over \sqrt{\Delta m^2 _{\rm sol}}}\right)$ for NH, and this is only warrantied in the hierarchical scenario where $m_{\rm light}$ is smaller than $0.05$eV. 

A possible extension of the minimal scenario that gives rise to the Weinberg operator is obviously the addition of extra Majorana singlets. In particular, in the case where one more neutrino is added, it is necessary that the mass is larger and/or the mixings smaller. This is for example the case in the $\nu$MSM \cite{Asaka:2005pn}, where the extra keV state is very weakly coupled, so that the contribution to the lightest neutrino mass is below $10^{-5}$eV, well below the above requirement. 

\begin{figure}
\begin{center}
 \includegraphics[width=0.7\columnwidth]{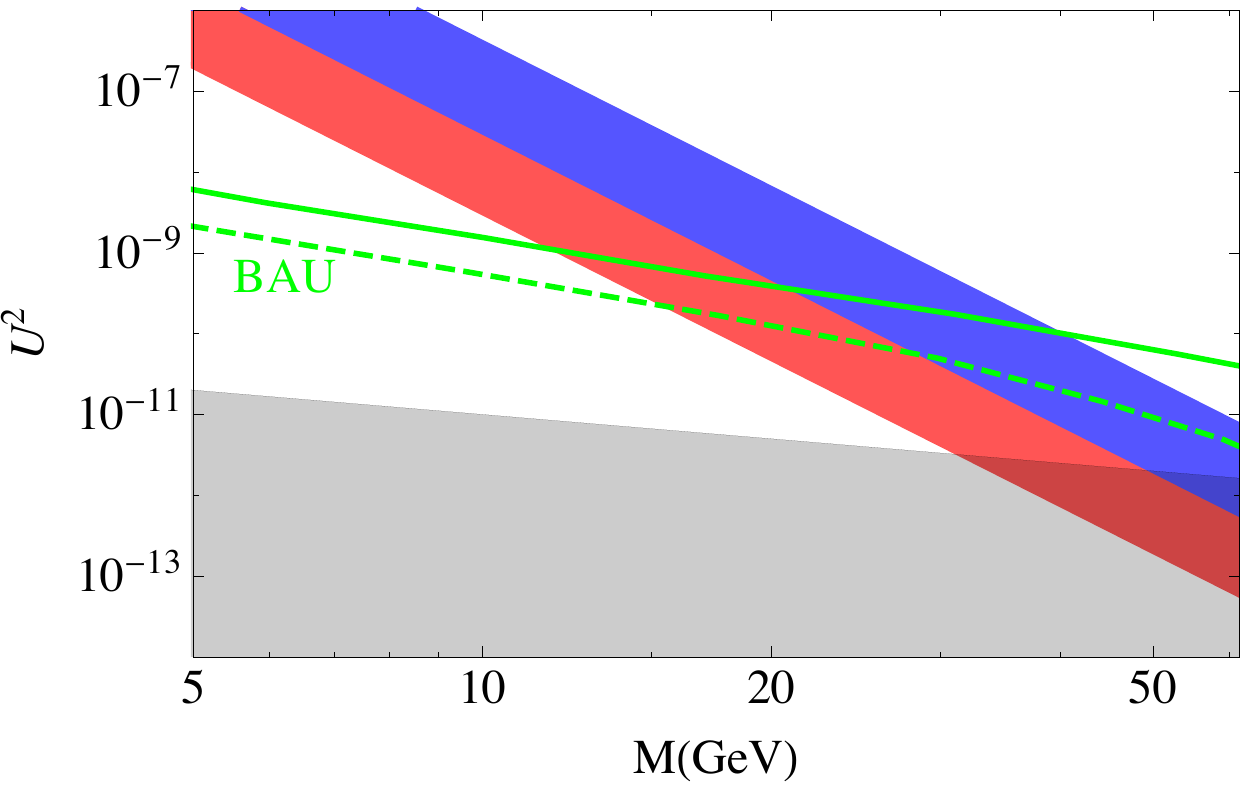} 
 \caption{Regions on the plane $(M, U^2)$  where LHC displaced track selection efficiency (eq.~(\ref{eq:tracker}) and (\ref{eq:chamber})) is above 10$\%$ in the IT (blue band) and MC  (red band). The grey shaded region cannot explain the light neutrino masses and the green lines correspond to the upper limits of the 90$\%$CL bayesian region for successful baryogenesis  in the minimal model for NH (solid) and IH (dashed), taken from \cite{Hernandez:2016kel}.}
 \label{fig:prop}
 \end{center}
\end{figure}

On the other hand, large hierarchies $\alpha_W \ll \alpha_{N\Phi} \sim \alpha_{B}$ could be present undisrupted by radiative corrections. In this case, direct bounds on the other two $d=5$ operators might be competitive and offer a new window into neutrino physics at the LHC. We have considered the bounds on $\alpha_{N\Phi}$ from 
searches of displaced leptons at LHC and we have found that LHC with 300 fb$^{-1}$ at 13TeV could set bounds
\begin{eqnarray}
\left|{\alpha_{N\Phi} v \over \sqrt{2} \Lambda}\right| \leq 10^{-3}-10^{-2} \rightarrow {\alpha_{N\Phi} \over \Lambda} \leq  6 \times (10^{-3}-10^{-2}) {\rm TeV}^{-1}.
\end{eqnarray}
It is important to note that if the coefficient of this operator is above this sensitivity limit, LHC could detect the sterile neutrinos for significantly smaller mixings than it is possible in the minimal model, in particular LHC could  even reach the seesaw limit as shown in Fig.~\ref{fig:prop}. 

The bounds on $\alpha_{NB}$ were considered in ref.~\cite{Aparici:2009fh} and found to be
\begin{eqnarray}
{\alpha_{NB} \over \Lambda} \leq  10^{-2} -10^{-1} {\rm TeV}^{-1}.
\end{eqnarray}
These operators could appear at tree level in extensions with scalar singlets, such as the Majoron model, where the singlet can couple to the singlet Majorana contraction $\bar{N} N^c$ and the Higgs portal $\Phi^\dagger \Phi$.  The exchange of the singlet scalar leads at tree level to the operator ${\mathcal O}_{N\Phi}$. On the other hand the operator ${\mathcal O}_{N B}$ needs to be generated at one loop. 

Finally flavour symmetries could also explain a large hierarchy of $\alpha_W$ and $\alpha_{N\Phi}$ or $\alpha_{N B}$. An approximate $U(1)_L$  could explain the hierarchy $\alpha_W \ll \alpha_{N\Phi}, \alpha_{NB}$. Such symmetry is also  the most natural scenario within the minimal model to have mixings significantly larger than the naive seesaw limit as required for 
their observability. Similarly, hierarchies of this type or $\alpha_W, \alpha_{NB} \ll \alpha_{N\Phi}$ are expected in the context of  minimal flavour violation \cite{Graesser:2007yj}. 

The presence of large contributions from these operators are not expected to modify the predictions concerning the flavour structure of the mixings of the heavy neutrinos, since the higher order corrections to the neutrino masses are very small, but they could provide a new portal at colliders to reveal the mechanism behind neutrino masses. On the other hand the same interactions will surely affect the  baryogenesis scenario with respect to the minimal model, since the sterile neutrinos can reach thermal equilibrium if the interactions induced by the higher dimensional operators are within LHC reach. The impact of ${\mathcal O}_{N\Phi}$ in the context of baryogenesis has been recenlty considered in \cite{Asaka:2017rdj}. 
 
\section*{Appendix A}

We list the relation between $\widetilde{m}$ and $\widetilde{U}$ and the physical neutrino masses, $m_i$, and mixings, at leading order in a perturbative expansion in $\delta m_\nu^W$:
\bea
\text{NH}&&
\nonumber\\
\widetilde{m_1}&=&0,
\nonumber\\
\widetilde{m_2}&=&m_2+\text{Re}\left(\frac{-e^{2 i \phi_1 }}{2}\left[2\delta_{11}s_{12}^2+
(\delta_{22}-2\delta_{23}+\delta_{33})c_{12}^2 +\sqrt{2}(\delta_{12}-\delta_{13})\sin2\theta_{12}\right]\right) +{\mathcal O}(\delta \epsilon),
\nonumber\\
\widetilde{m_3}&=&m_3+\text{Re}\left(\frac{-e^{2 i \phi_1}}{2}\left[\delta_{22}
+2 \delta_{23}+\delta_{33}\right]\right) +{\mathcal O}(\delta \epsilon),
\nonumber\\
\text{IH}&&
\nonumber\\
\widetilde{m_1}&=&m_1+\text{Re}\left(\frac{1}{2}\left[2\delta_{11} c_{12}^2+ (\delta_{22}
-2 \delta_{23}+\delta_{33}) s_{12}^2 +\sqrt{2}(\delta_{12}-\delta_{13})\sin2\theta_{12}\right]\right) +{\mathcal O}(\delta \epsilon),
\nonumber\\
\widetilde{m_2}&=&m_2+\text{Re}\left(\frac{-e^{2i \phi_1 }}{2}\left[2\delta_{11}s_{12}^2+
(\delta_{22}-2\delta_{23}+\delta_{33})c_{12}^2 +\sqrt{2}(\delta_{12}-\delta_{13})\sin2\theta_{12}\right]\right)+{\mathcal O}(\delta \epsilon),
\nonumber\\
\widetilde{m_3}&=&0.
\eea
the next to leading order contributions are $\mathcal{O}\left(\epsilon\, \delta m \right)$.

Defining %
\be 
\widetilde{U}=U (1 + \delta \tilde U),
\ee
at leading order we find
\be
\delta \tilde U =
\begin{pmatrix}
0 & \delta \tilde U_{12} & \delta \tilde U_{13} \\
-\delta \tilde U_{12}^* &  0 & \delta \tilde U_{23} \\
-\delta \tilde U_{13}^* & -\delta \tilde U_{23}^* &0
\end{pmatrix} + {\mathcal O}(\delta^2),
\ee
with 
\bea
\text{NH}&&
\nonumber\\
\delta \tilde U_{12}&=&\frac{e^{-i(\phi_1+\phi_2)}}{2r\sqrt{\Delta m^2_{atm}}}\left[
2\sqrt{2}(\delta_{12}-\delta_{13})\cos2\theta_{12}+
(2\delta_{11}+2\delta_{23}-\delta_{22}-\delta_{33})\sin2\theta_{12}\right],
\nonumber\\
&+& \mathcal{O}\left( \delta  \right)
\nonumber\\
\delta \tilde U_{13}&=&\frac{e^{-i\phi_2}}{2\sqrt{\Delta m^2_{atm}}}
\left[ \sqrt{2}(\delta_{12}+\delta_{13})c_{12} +(\delta_{33}-\delta_{22})s_{12}\right]
+\mathcal{O}\left(\epsilon\, \delta  \right),
\nonumber\\
\delta \tilde U_{23}&=&\frac{e^{-i\phi_1}}{2\sqrt{\Delta m^2_{atm}}}
\left[-\sqrt{2}(\delta_{12}+\delta_{13})s_{12} +(\delta_{33}-\delta_{22})c_{12}\right]
+\mathcal{O}\left(\epsilon\, \delta  \right),
\nonumber\\
\text{IH}&&
\nonumber\\
\delta \tilde U_{12}&=&\text{Re}\left(\frac{e^{-i\phi_1}}{2r^2\sqrt{\Delta m^2_{atm}}}\left[
2\sqrt{2}(\delta_{12}-\delta_{13})\cos2\theta_{12}-
(2\delta_{11}+2\delta_{23}-\delta_{22}-\delta_{33})\sin2\theta_{12}\right]\right),
\nonumber\\
&+& \mathcal{O}\left( \delta  \right)
\nonumber\\
\delta \tilde U_{13}&=&\frac{e^{i\phi_2}}{2\sqrt{\Delta m^2_{atm}}}
\left\{(\delta_{22}^*-\delta_{33}^*)s_{12} -\sqrt{2}(\delta_{12}^*+\delta_{13}^*)c_{12} \right\}
+\mathcal{O}\left(\epsilon\, \delta  \right),
\nonumber\\
\delta \tilde U_{23}&=&\frac{e^{i(\phi_1+\phi_2)}}{2\sqrt{\Delta m^2_{atm}}}
\left\{(\delta_{33}^*-\delta_{22}^*)c_{12}-\sqrt{2}(\delta_{12}^*+\delta_{13}^*)s_{12} \right\}
+\mathcal{O}\left(\epsilon\, \delta  \right).
\nonumber\\
\eea
\begin{acknowledgments}

We  thank P.~Coloma, L.~Delle Rose, E.~Fern\'andez-Mart\'{\i}nez, O.~Mattelaer, R.~Ruiz and T. Schwetz for useful discussions and/or clarifications. 
This work was partially supported by grants FPA2014-57816-P, PROMETEOII/2014/050, and  the European projects 
H2020-MSCA-ITN-2015//674896-ELUSIVES and 690575-InvisiblesPlus-H2020-MSCA-RISE-2015. 

\end{acknowledgments}

\bibliographystyle{JHEP}
\bibliography{biblio}

\end{document}